# Enzymatic AND Logic Gates Operated Under Conditions Characteristic of Biomedical Applications

Dmitriy Melnikov,[a] Guinevere Strack,[b] Jian Zhou,[b] Joshua Ray Windmiller,[c] Jan Halámek,[b] Vera Bocharova,[b] Min-Chieh Chuang,[c] Padmanabhan Santhosh,[c] Vladimir Privman,[a]* Joseph Wang,[c]** and Evgeny Katz[b]***

[a] *Department of Physics, Clarkson University, Potsdam, NY 13676*
[b] *Department of Chemistry and Biomolecular Science, Clarkson University, Potsdam, NY 13676*
[c] *Department of NanoEngineering, University of California – San Diego, La Jolla, CA 92093*

---

*E-mail: privman@clarkson.edu; Tel.: +1 (315) 268-3891; Fax: +1 (315) 268-6610
**E-mail: josephwang@ucsd.edu; Tel.: +1 (858) 246 0128; Fax: +1 (858) 534 9553
***E-mail: ekatz@clarkson.edu; Tel.: +1 (315) 268-4421; Fax: +1 (315) 268-6610

---

## Abstract

Experimental and theoretical analyses of the lactate dehydrogenase and glutathione reductase based enzymatic **AND** logic gates in which the enzymes and their substrates serve as logic inputs are performed. These two systems are examples of the novel, previously unexplored, class of biochemical logic gates that illustrate potential biomedical applications of biochemical logic. They are characterized by input concentrations at logic **0** and **1** states corresponding to normal and abnormal physiological conditions. Our analysis shows that the logic gates under investigation have similar noise characteristics. Both significantly amplify random noise present in inputs, however we establish that for realistic widths of the input noise distributions, it is still possible to differentiate between the logic **0** and **1** states of the output. This indicates that reliable detection of abnormal biomedical conditions is indeed possible with such enzyme-logic systems.

---

Web-link to future updates of this article: www.clarkson.edu/Privman/229.pdf

## 1. Introduction

Recent developments in the areas of chemical[1] and biomolecular[2] computing, emerging from the broader framework of unconventional computing,[3] have resulted in the realization of various biochemical Boolean logic gates[4-6] and some other building blocks of digital logic, as well as small networks.[7] Functional units demonstrated, have included molecular memory units,[8] comparators,[9] multiplexers/demultiplexers,[10] encoders/decoders,[11] etc., which are expected to be used as components for assembling "devices" processing information by chemical means. These systems are capable of performing simple arithmetic functions, for instance, operating as half-adder/half-subtracter[12] or full-adder/full-subtracter.[13] Chemical systems can mimic keypad lock devices[14] and even operate as molecular automata[15] based on networked logic operations.

Despite the fact that chemical computing is a rapidly developing area of research, most of the reported studies have focused on the demonstration of basic concepts, while aspects of practical applications have not been clearly addressed. In order to justify the studies, conjectures regarding possible relevance to molecular computer designs have been routinely put forth. In reality, however, present-day chemical systems are not able to directly compete with electronic computers. Indeed, biocomputing systems have thus far been connected only in small "circuits" capable of carrying out basic arithmetic operations on the time-scale of minutes or longer. On the other hand, application of biomolecular logic systems for analytical purposes could yield an entirely new functionality: a novel class of bio(chemical) sensors which are able to accept many input signals and produce binary alert-type outputs in the form of "YES"-"NO" to identify relevant biomedical conditions.[16] This approach has already been successfully applied to analyze multi-parameter physiological conditions corresponding to different kinds of injuries.[17] The biosensors being developed, have been based on enzyme systems logically processing different biomarker signals changing from normal physiological concentrations to elevated (or eventually decreased) levels as a result of injuries, and will require careful consideration of noise control in gate and network implementation.[18-21]

In the present study experimental and theoretical analyses of two kinetically similar **AND** logic gates activated by enzymes and their corresponding substrates were performed. The first

**AND** logic gate was activated by lactate dehydrogenase (LDH) and lactate — jointly constituting a definitive set of biomarkers for abdominal trauma (ABT),[22] while the second **AND** logic gate was activated by glutathione reductase (GR) and glutathione disulfide (GSSG), which are indicative of oxidative stress when their concentrations are elevated.[23]

It should be noted that most of the previously developed (bio)chemical computing systems have utilized arbitrary levels of input signals (usually logic **0** corresponded to physical zero concentration of inputs, while logic **1** was selected at a convenient concentration of reacting species).[1,4-7,18-20] Novel biomedical applications require the operation of biochemical digital computing systems at specific levels of logic inputs determined by physiological conditions with **0** and **1** corresponding to "normal" and "abnormal" concentrations. The logic **0** and **1** input signals may accordingly have small range of variation, thus resulting in poor "digital" discrimination of the generated output signals. Furthermore, the output signal values interpreted as logic **0** and **1** can have significant variation as compared to the difference between the two reference logic points. Finally, functioning at physiological concentrations imposes strict constrains on the possible values or ranges of concentrations, and leaves little margin for variation of the process parameters for optimizing the gate functioning. In order to consider the quality and possible optimization of the biochemical logic gates operation, the whole surface-response function should be evaluated and analyzed for variable concentrations of input signals.[18-20]

Previously implemented biochemical logic gates typically utilized enzymes operating as the gate's "machinery," which served to process signals represented by the corresponding substrates/cosubstrates.[5] Alternatively, enzymes were used as input signals to activate the "machinery-soup" composed of all the other reactants.[6,18] The present biocomputing gates are activated upon the simultaneous supply of an enzyme and a corresponding substrate, as inputs for the **AND** logic operation. This poses additional challenges for modeling of system performance and from this point of view such gates can be regarded as examples of a novel, previously unexplored, class of biochemical logic systems. Since input signals in these gates correspond to physiological conditions, there is always some uncertainty present in their values, because chemical concentrations obviously vary from case to case, i.e., there is some distribution of input signals or, in other words, noise. As our systems are illustrative of a broad range of biomedical applications, analysis and understanding of how this noise in the input propagates in such logic gates are extremely important.

This paper is organized as follows. In Section 2, we provide details of the experimental procedure. Then in Section 3, we discuss the model utilized to analyze noise characteristics of these two gates. Analysis of the experimental data and numerically computed response surfaces and the resulting noise-amplification features of the two gates are reported in Section 4. Finally, the results are summarized in Section 5.

## 2. Experimental Section

*Chemicals and Materials.* The enzymes and other chemicals were obtained from Sigma-Aldrich and were used without further purification: glutathione reductase (GR, from *S. cerevisiae*, E.C. 1.6.4.2), lactate dehydrogenase (LDH, from porcine heart, E.C. 1.1.1.27), β-nicotinamide adenine dinucleotide dipotassium salt (NAD$^+$), β-nicotinamide adenine dinucleotide 2'-phosphate reduced tetrasodium salt (NADPH), L(+)-lactic acid, L-glutathione disulfide (GSSG), cobalt(II) phthalocyanine (CoPC), methylene green (MG), dithio*bis*-(2-nitrobenzoic acid) (DTNB – Ellman's reagent) and other standard inorganic salts/reagents. Ultrapure water (18.2 MΩ cm) from NANOpure Diamond (Barnstead) source was used in all of the experiments.

*Instrumentation and measurements.* A CH Instruments model 1232A potentiostat was used for all the electrochemical measurements and a Shimadzu UV-2450 UV-Vis spectrophotometer (with a TCC-240A temperature-controlled cuvette holder and 1 mL PMMA cuvettes) was used for all the optical measurements. A Mettler Toledo SevenEasy s20 pH-meter was employed for pH measurements. A Barnstead Thermodyne Cimarec stir / heat plate was employed to continuously agitate solutions and maintain temperature at 37°C while electrochemical measurements were performed. All optical measurements were performed in temperature-controlled cuvettes/cells at 37°C mimicking physiological conditions and all reagents were incubated at this temperature prior to experimentation. All electrochemical measurements were performed on a screen-printed electrode (SPE) and the potentials were measured vs. a quasi-reference Ag/AgCl electrode.

*Screen-Printed Electrode (SPE) Preparation:* A screen printed three-electrode strip, custom-designed using AutoCAD®, consisted of a circular carbon working electrode (geometrical area: 3 mm$^2$) inscribed in hemispherical counter (area: 10 mm$^2$), and reference electrodes (area: 2 mm$^2$). The fabrication of the flexible screen-printed electrode system is detailed as follows: An Ag/AgCl-based ink from Ercon (E2414) was employed to define the conductive underlayer as well as the reference electrode. A carbon-based ink (Ercon E3449) was subsequently overlaid on the conductive underlayer to define the counter and working electrode geometries. In the experiments with GR/GSSG, CoPC was dispersed in the carbon ink (2% w/w). Finally, an insulator ink (Ercon E6165) was overlaid on the Ag/AgCl and carbon layers to insulate all except the contact pads and the upper active segment of the electrodes. A Speedline Technologies MPM-SPM screen printer was used to print the pattern onto a 250 μm-thick flexible polyethylene terephthalate substrate (DuPont Melinex 329). Subsequent to the printing process, the patterned substrate was cured in a temperature-controlled convection oven (SalvisLab Thermocenter) at 120°C for 20 min. The substrate was finally cleaved to create single-use test strips possessing overall dimensions of 10 mm × 34 mm.

*Experimental Procedure for the lactate / LDH **AND** Gate:* A graphical representation of the enzymatic **AND** logic gate is outlined in Scheme 1. The gate "machinery" consisted of NAD$^+$ (10 mM and 1 mM for optical and electrochemical measurements, respectively) in 50 mM sodium/potassium phosphate buffer with 0.2 mM MgCl$_2$ and 0.01 mM CaCl$_2$, pH = 7.15. MG, 1 mM, was added as a mediator catalyzing NADH oxidation for the electrochemical measurements. The **AND** logic gate was activated by lactate and LDH as **Input 1** and **Input 2**, respectively. Logic **0** and **1** levels of lactate (1.6 and 6.0 mM) and LDH (150 and 1000 UL$^{-1}$ — given in activity units per liter) input signals were chosen according to the mean normal and elevated physiological concentrations of these biomarkers relevant for the diagnosis of abdominal trauma.[22] The optical measurements employed the transmission method (absorbance at $\lambda$ = 340 nm) and the electrochemical measurements were performed using the amperometric technique (continuous agitation and 0.1 V applied at the working electrode vs. an Ag/AgCl reference) in order to monitor the generation of the NADH product. In the optical experiments the absorbance measurements were started immediately after mixing the reagents in a cuvette and the final absorbance was taken at 360 sec from the beginning of the measurements. In the electrochemical experiments, the buffer, lactate, LDH, and MG solutions were dispensed in the microcell and the amperometric recording initiated. Following a 150 sec settling period to allow the background current to decay to stable levels, the $NAD^+$ solution was added and the recording continued for an additional 150 sec. At 300 sec following the initiation of the recording (150 sec following the addition of $NAD^+$), the output-signal current reading was extracted. In order to remove the contribution of the background current to the bona-fide electrochemical signal arising from the biocatalytic process, the current obtained just prior to the addition of $NAD^+$ at 150 sec was subtracted from the current reading at 300 sec. In order to map the response-surface of the **AND** gate, we varied lactate and LDH concentrations obtaining an array of 6×6 experimental points [concentration step in electrochemical (optical) measurements was 1.5 (1.2) mM, with variation between 0 (1) and 7.5 (7) mM for lactate, and step of 250 (200) $UL^{-1}$, with variation between 0 (100) and 1250 (1100) $UL^{-1}$ for LDH]. Note that the varied concentrations of the input signals scanned the ranges somewhat below the logic **0** values to somewhat above the logic **1** values.

*Experimental Procedure for the GSSG / GR AND Gate:* A graphical representation of the enzymatic **AND** logic gate is outlined in Scheme 2. The gate "machinery" consisted of 0.18 mM NADPH (DTNB, 2 mM, was added for optical measurements) in 50 mM citrate buffer, pH = 5.0. The **AND** logic gate was activated by GSSG and GR as **Input 1** and **Input 2**, respectively. Logic **0** and **1** levels of GSSG (150 and 400 μM) and GR (556 and 650 $UL^{-1}$) input signals were chosen according to the mean normal and elevated physiological concentrations of these biomarkers in erythrocytes, motivated by studies of oxidative stress.[23] Since GR and GSSG are mainly present in intracellular compartments of erythrocytes, maintaining a physiological pH (7.35-7.45 for serum) was not essential for the assay. Thus, in order to attenuate the enzymatic reaction and to provide output signals significantly different for each level of inputs, the pH was experimentally optimized and adjusted to the final value of 5. In the optical experiments the thiol groups of the biocatalytically produced GSH irreversibly reacted with DTNB resulting in the formation of thio-(2-nitrobenzoic acid), TNB, which was monitored by absorbance changes at $\lambda = 412$ nm. The absorbance changes were measured during the reaction following the reactants mixing in a cuvette and the final absorbance value was taken at 100 sec from the beginning of the reaction. In the electrochemical experiments CoPC (2% w/w) was employed as a redox mediator to enable the low-potential detection of GSH at 0.5 V vs. Ag/AgCl reference. The buffer, GR, GSSG, and NADPH solutions were dispensed on the CoPC-modified SPE and the chronoamperometric recording (in a quiescent solution) initiated. Following a 30 sec settling period to allow the transient current to decay to negligible levels, the recording was terminated and the reading was extracted. In order to map the response-surface of the **AND** gate, we varied the input concentrations and obtained an array of 6×6 experimental points (concentration step of 83 µM between 67 and 483 µM for GSSG and concentration step of 31 UL$^{-1}$ between 524 and 679 UL$^{-1}$ for GR). Note that the varied concentrations of the input signals here also started below the logic-**0** values, and then again exceeded the logic-**1** values.

## 3. Model of the AND Gate Function

As mentioned in Section 2, the initial concentrations of **Input 1** and **Input 2** chemicals vary between some minimum (not equal to zero), $[\textbf{Input 1}](t=0) = C_{1,\min}$, $[\textbf{Input 2}](t=0) = C_{2,\min}$, and maximum, $[\textbf{Input 1}](t=0) = C_{1,\max}$, $[\textbf{Input 2}](t=0) = C_{2,\max}$, values determined by specific biomedical applications. The output product concentration, $P$, is measured as $P(t=t_{\text{gate}})$ at specific reaction time, $t_{\text{gate}}$, and also varies between two values, $P_{\min}$ and $P_{\max}$, as the input concentrations are swept from their minimal to maximal values. To analyze our logic gates, we cast input/output chemical signals in terms of the dimensionless input $(x, y)$ and output $z$ variables scaled to the "logic" ranges,

$$x = \frac{C_1 - C_{1,\min}}{C_{1,\max} - C_{1,\min}}, \quad y = \frac{C_2 - C_{2,\min}}{C_{2,\max} - C_{2,\min}}, \quad z = \frac{P - P_{\min}}{P_{\max} - P_{\min}}. \tag{1}$$

The noise amplification properties of the **AND** gate are conveniently analyzed[18-21,24] by considering the function $z(x,y)$, termed the response surface, in the vicinity of the logic points $(x,y) = (\mathbf{0,0}), (\mathbf{0,1}), (\mathbf{1,0}), (\mathbf{1,1})$. In general, the gate output depends not only on the initial inputs and the time of the reaction, but also on other system parameters such as, for example, in the present case the initial concentration of the cosubstrate, $c$,

$$P = F(C_1, C_2; c, \text{pH}, ...; t_{\text{gate}}, ...). \tag{2}$$

Note that we divided these additional parameters, which could be physical or (bio)chemical, in Equation (2) into two groups. Certain parameters, here exemplified by $c$ and pH, can be adjusted to some extent to improve the gate performance. Other parameters cannot be easily adjusted, here, for instance, $t_{\text{gate}}$, which was chosen to have the best possible separation between the **1** and **0**s of the output, as well as most physical properties such as the temperature, which are fixed by the intended application.

In order to calculate the function $z(x,y)$, we need to model the parameter dependence in Equation (2). It is important to note[24] that there are several sources of noise to be considered in biochemical computing. In addition to the natural fluctuations in the inputs about the precise logic-point values, there can also be systematic deviations, as well as uncertainty in the physiological concentrations which replace the sharp logic-value definitions with ranges. The output, similarly, is not precisely defined. In addition to the spread of their values, outputs can be somewhat shifted from the selected reference "logic" answers. Furthermore, the shifts in the outputs will generally be somewhat different at different logic inputs which are supposed to yield the same logic-value answers in the binary convention: here the three inputs $(\mathbf{0,0}), (\mathbf{0,1}), (\mathbf{1,0})$, which are all expected to yield **0** output.

Thus, the functional dependence in Equation (2) might not be precisely defined, and should be viewed as an average with possible small systematic shifts from the "ideal" logic values built in. One way to evaluate the function *F* by fitting data from experiments, is to solve a set of precise or phenomenological kinetic equations corresponding to enzymatic reactions; this was the approach taken in previous works[18-20] for several biochemical systems. Rate constants of the involved chemical reactions were then treated as adjustable parameters with values determined from the numerical fitting of the experimental data to the solution of the kinetic equations. This approach requires a working knowledge for the kinetics of the studied logic gate, which is not the case for the present systems due to their complexity. Indeed, even a simplified kinetic description[25] would require fitting too many adjustable parameters which consequently cannot be accurately determined from the available gate-response mapping data.

Another approach[21] has been to specify a phenomenological fit function for *F*, based on our experience with the solution of the kinetic equations as well as on the experimental expectations and observations. This approach has been developed[21,24] for general analysis of the noise performance of logic gates and networks, when detailed information of the system's kinetics is either unavailable or the set of the reactions is far too complex to determine fitting parameters uniquely.

Indeed, we are only interested in the global features of the response surface function $z(x,y)$, sufficient to evaluate its behavior in the vicinity of the four logic-point values. However, an earlier phenomenological fitting form[21] developed for typical biochemical reactions encountered in **AND**-gate realizations with the logic **0**s at zero concentrations, actually reflects the features of the biochemical reaction rather than more generally those of the binary logic. Therefore, it is inappropriate to use such a form for gates with non-zero logic-**0** values. To avoid this complication, we did not rescale all the data upfront to the logic ranges, according to Equation (1). Instead, the idea is to fit the function *F* in Equation (2) first, by using the phenomenological shape-function suggested[21] by earlier studies, which here can be written as

$$F = \frac{\phi C_1 C_2}{(\alpha C_1 + 1)(\beta C_2 + 1)}. \tag{3}$$

The resulting function will then be used to derive an estimate for $z(x,y)$. We point out, however, that the overall constant $\phi$ in Equation (3) is not a useful fitting parameter because this factor will cancel in the calculation of $z(x,y)$. Therefore, the actual data fits, described below, were done for *F* normalized to (divided by) its measured value at input concentrations at which it was the largest, and this ratio was least-squares-fitted in order to determine the two fitting constants of interest, $\alpha$ and $\beta$.

Because of the non-zero values of the minimum input concentrations, the output concentration possesses somewhat different values for the combinations of the inputs $(C_{1,\min}, C_{2,\min})$, $(C_{1,\min}, C_{2,\max})$, $(C_{1,\max}, C_{2,\min})$. In other words, output at logic **0** has three hopefully close but different values. However, for a "standard" **AND** gate, the logic-**0** output should be equal for all these inputs. This implies that enzymatic gates considered in this work have some a systematic "noise" built-in.[24] This "noise" exists in addition to the usual, random noise which was extensively analyzed earlier.[18-20] Thus, the demands for additional network elements with filtering properties,[18,24] will be even more stringent here if the present systems are to be used as components of biochemical logic networks. We note that the present modeling approach is also somewhat different from earlier variants[20] in that the mean values to be used (defined) as $P_{\min}$ and $P_{\max}$ are not known upfront. Rather, we first evaluate the function in Equation (2) and then calculate $P_{\max}$ and the smallest logic **0** value as $P_{\min}$ (another option would be, e.g., an average of the three logic **0** outputs as $P_{\min}$). Indeed, the output signals to be used as reference **0** and **1**, are not arbitrary but are established by the application, up to the aforementioned built-in noise, which should be small to have a high-quality realization of the **AND** function.

If the phenomenological fit of Equation (3) is a good one, then $\alpha(c, \text{pH},...; t_{\text{gate}},...)$ and $\beta(c, \text{pH},...; t_{\text{gate}},...)$ should be functions of the chemical and physical parameters introduced in Equation (2). The present approach, however, does not attempt to obtain this kinetic information, and therefore offers no quantitative information on the dependence of the response surface function $z$ on such parameters (in addition to its arguments $x$ and $y$). Thus the present approach cannot be used to directly optimize gate functionality. However, qualitative arguments[20,24] can usually be utilized to decide whether to increase or decrease of the overall gate activity to improve its performance. These usually require large changes in the "gate machinery" properties because the response surface is obtained in terms of the scaled variables, Equation (1), and is not sensitive to the leading order linear-response-type changes. For the present systems, with the chemical concentrations established at physiological conditions, large changes are not easy to achieve, and therefore the gate performance per se cannot be readily optimized by varying the few experimental variables under our control. Instead, we aim at estimating the degree of "noisiness" of the realized gates' operation, in expectation that networking with other elements (reaction steps) such as filters, when developed, will be a proper approach to obtain systems with less noise.

To analyze random noise amplification from input to output, we calculate[18,19] the noise amplification factors as the ratio of the output noise distribution spreads, $\sigma_{ij}^{out}$, to the fixed input noise distribution width, $\sigma^{in}$, with the definition

$$\sigma_{ij}^{out} = [\langle z^2 \rangle_{ij} - \langle z \rangle_{ij}^2]^{1/2}, \tag{4}$$

where the averages $\langle \cdots \rangle$ at each logic point, $(i, j) = \mathbf{(0,0)}, \mathbf{(0,1)}, \mathbf{(1,0)}, \mathbf{(1,1)}$, are computed with respect to the input noise distribution $D_{ij}(x, y) = X_i(x) Y_j(y)$, which for simplicity is assumed to be uncorrelated: a product of Gaussians with equal width, $\sigma^{in}$, in terms of the scaled variables. We used the most straightforward expression,

$$\langle z^n \rangle_{ij} = \iint z^n(x, y) D_{ij}(x, y) dx dy, \tag{5}$$

for numerical computations. For the description of the effect of the systematic noise — the built-in shift which translates into imprecise average values for $z$ — we first compute the averages, $\langle z \rangle_{ij}$, and then the spread interval $(\langle z \rangle_{ij} - \sigma_{ij}^{out}, \langle z \rangle_{ij} + \sigma_{ij}^{out})$ which defines the region where the output corresponding to the specific logic point is most likely to be found for the assumed distribution of random noise at the input. If the combined spread region for the three logic-**0** points overlaps with or is too close to that for the logic-**1** point, then the present gates cannot be used for systems with the degree of random noise in the input at the level of the $\sigma^{in}$ value considered.

## 4. Results and Discussion

### 4.1. *Lactate/LDH AND Gate*

Catalytic reduction of $NAD^+$ to NADH proceeding in the presence of lactate (**Input 1**) and LDH (**Input 2**) results in the increased concentration of NADH in the solution which can be followed by optical and electrochemical means; see Scheme 1. In the optical measurements the absorbance increase characteristic of NADH formation was monitored at λ = 340 nm (note that the system's background absorbance was subtracted), Figure 1(A). In the electrochemical experiments, the NADH oxidation was mediated by MG,[26] and the obtained current values corresponded to the concentration of NADH produced in the course of the biocatalytic reaction, Figure 1(B). The system represented **AND** logic gate when the reaction results in the product formation only in the presence of the both reacting species: the substrate (lactate) and the enzyme (LDH) [input signal combination (**1,1**)], Figure 1(C). However, since the logic **0** values of both input signals do not correspond to zero concentrations of the reactants, the reaction product is also generated at **0** logic values of one or both inputs [input combinations (**0,0**), (**0,1**), (**1,0**)]. In order to analyze the output function of the biocatalytic system for the given logic values of the input signals we performed the measurements with several varied concentrations of the substrate (lactate) and the enzyme (LDH).

The normalized (to the maximum observed value) experimental response surface obtained by optical absorbance spectroscopy (top row) and electrochemical current measurements (bottom row) is shown in Figure 2 for a reaction time $t_{gate} = 360$ sec and 300 sec, respectively. As one can see, both the measured [Figs. 2(A) and 2(D)] and the fitted [Figs. 2(B) and 2(E)] surfaces obtained by the two different methods are quite consistent. The logic surfaces extracted from optical and electrochemical measurements, shown Figures 2(C) and 2(F), are very similar to one another. This is also confirmed by comparing the output logic values at four logic points from the two data sets, see Figure 3. In this plot, the *z*-values corresponding to (**0,0**) and (**1,1**) logic input combinations are at zero and one, respectively, while *z*-values for two other logic input pairs, (**0,1**) and (**1,0**), fall in between. Note that the outputs from the latter pair are not (close to) zero as they should be for a true **AND** gate because, as already mentioned, they correspond to different nonzero concentrations of input chemicals and, as such, cannot be made equal to each other or to the output at **(0,0)**.

The similarity between the two response surfaces is also manifested in similar noise characteristics evaluated for each response. Both the noise amplification factors, illustrated in Figures 4(A) and 4(C), i.e., the degree to which the random input noise is amplified at a particular logic point, and the spread regions, shown in Figures 4(B) and 4(D), that provide information on the separation between the logic points for a given value of the input noise distribution width, agree both qualitatively and quantitatively. From the plots of $\sigma_{ij}^{\text{out}}/\sigma^{\text{in}}$, we can gauge the response of the **AND** gate to noise in its inputs: If the maximum of $\sigma_{ij}^{\text{out}}/\sigma^{\text{in}}$ is larger/smaller than one, then the logic gate amplifies/suppresses incoming random noise. In particular, one can see that the "worst" (most noisy) logic point irrespective of the input noise spread, that with the largest ratio $\sigma_{ij}^{\text{out}}/\sigma^{\text{in}}$, is **(1,0)**, at which the lactate/LDH concentration is the largest/smallest. Even though the noise is actually suppressed at the **(1,1)** logic point ($\sigma_{ij}^{\text{out}}/\sigma^{\text{in}} \approx 0.5 < 1$), poor performance ($\sigma_{ij}^{\text{out}}/\sigma^{\text{in}} > 2$, see Figure 4) at the **(1,0)** logic point makes this system in general not very suitable for incorporation in large networks of biochemical logic gates without additional filtering elements[18,24] aimed at reducing noise.

The present approach of estimating the random noise amplification factor by assuming a Gaussian input distribution, becomes inappropriate for $\sigma^{\text{in}}$ approximately exceeding 0.2, when the spread of the distribution becomes comparable to a sizable fraction of the unit interval. While the $\sigma_{ij}^{\text{out}}/\sigma^{\text{in}}$ curves tend to decrease for $\sigma^{\text{in}}$ beyond ~0.1, the overall shape of $z(x,y)$ in our systems is smooth-convex, which means that the **AND** gate always amplifies analog noise.[18]

In practice, the physiological spread of the input concentrations is much less than 20%, at least for the **(0,0)** logic point. Indeed, the observed concentrations for lactate are within the 0.5-2.25 mM range, while the LDH values are spread over the 42-180 U/L interval.[27] The average values that we employ as the input logic **0**s are 1.6 mM and 150 U/L, respectively. This indicates that the distribution of the inputs is not symmetrical, and that the distribution spread is different for lactate and LDH. Because of this, our calculations only yield a qualitative estimate of the noise amplification at this logic point: By taking $\sigma^{in} = 10\%$, which is the average of the data spreads for the *x* and *y* logic inputs (~15% and ~5%, respectively), we can deduce from Figures 4(A) and (C), that the noise will be amplified by the factor of ~120% at this logic point. Assuming that the input noise spread stays the same for all other logic points (for which we are not aware of results reported in the literature for the spreads of values), we can see that maximum noise amplification produced by this gate would be ~200%, at the **(1,0)** point.

This degree of amplification of the incoming noise does not, however, preclude practical utilization of the gate. Analysis of the spread region plots in Figures 4(B) and 4(D), shows that at $\sigma^{in} = 10\%$, the **(1,1)** logic point is well separated from other logic points by a "gap" of about 0.5, even though average outputs at **(0,1)** and **(1,0)** are not zero. However, if $\sigma^{in}$ increases to ~30%, the spread regions of **(1,1)** and **(1,0)** points begin to overlap, and it may not be possible to distinguish between the **0** and **1** levels of the output signal.

### 4.2. GSSG/GR AND Gate

Similarly to the previous system, the catalytic reduction of GSSG (**Input 1**) to GSH by GR (**Input 2**), that proceeds in the presence of NADPH (a part of the gate "machinery"), results in an elevated concentration of GSH in the solution which can be monitored by optical and electrochemical means, Scheme 2. For optical measurements, in order to convert GSH to a chromogenic product, DTNB (Ellman's reagent) was employed which resulted the formation of thio-(2-nitrobenzoic acid), TNB — monitored optically at $\lambda$ = 412 nm, Figure 5(A). For electrochemical experiments, the GSH oxidation was catalyzed by CoPC impregnated at an SPE.[28] The current obtained corresponded to the concentration of GSH produced in the course of the biocatalytic reaction, Figure 5(B). The system represents an **AND** logic gate when the reaction results in the formation a product only in the presence of both reacting species: the substrate (GSSG) and the enzyme (GR) — input signal combination **(1,1)**, Figure 5(C). However, analogous to the lactate/LDH system, logic **0** values of the input signals do not correspond to zero concentrations of the reactants, so that some amount of the reaction product is also generated at logic **0** values of one or both inputs. In order to analyze the output function of the biocatalytic system for the given logic values of the input signals, we performed the measurements for the variable concentrations of both logic inputs: the substrate (GSSG) and enzyme (GR). Unlike the lactate/LDH system, which can serve directly as a logic gate injury detecting system for abdominal trauma,[22] GR/GSSG is an example of a system which shows potential for intracellular investigations. Concentrations of the enzyme (GR) and its substrate (GSSG) used in the present study correspond to those inside erythrocytes. However, the presence of these biomarkers in blood is obvious when erythrocytes are ruptured, which can be symptom of radiation exposure or severe oxidative stress.[29]

In general, we found that the optically measured response surface and noise properties of this logic system were very similar to the one discussed in the previous subsection. One can see from the response surface shown in Figure 6 and normalized logic outputs in Figure 7 that the **(0,1)** and **(1,0)** points are displaced from zero, in direct correspondence with the results shown in Figures 2 and 3 for the other system. Also, the agreement between optical and electrochemical measurements at the logic points is satisfactory, as evident from Figure 7. Therefore, for the GR/GSSG system electrochemical measurements were performed only at the four logic points rather than mapping the entire response surface.

The maximum noise amplification factor here is also larger than 100%, see Figure 8(A), which is not surprising given the overall convex shape of the response surface in Figure 6. The computed ratios $\sigma_{ij}^{out}/\sigma^{in}$ are actually smaller than those calculated for the LDH system (see Figure 4). This indicates that the system exhibits a somewhat better noise performance, that is, it amplifies input noise to a lesser extent that the lactate/LDH logic gate (the maximum amplification factor here is 1.7 vs. 2.3 for the LDH based gate). However, because of the larger variation in the average *z*-values at the logic **0** points (see Figure 7), the spread-region gap separating the logic **1** from logic **0** outputs is actually smaller (~0.3, vs. ~0.5 for the lactate/LDH gate), as seen in Figure 8(B). This is because of a very small physiological range of GR concentration that effectively brings **0** and **1** values of the **Input 2** very close together.

## 5. Conclusion

In this work we performed experimental and theoretical analysis of two **AND** logic gates activated by enzymes and their corresponding substrates. The first **AND** logic gate was activated by lactate dehydrogenase (LDH) and lactate, while the second **AND** logic gate was activated by glutathione reductase (GR) and glutathione disulfide (GSSG). Logic **0** and **1** levels of input signals were chosen according to the normal and pathological concentrations of these biomarkers relevant to the diagnosis of abdominal trauma for the lactate/LDH system and oxidative stress for GSSG/GR system. In order to analyze the output function of the biocatalytic systems for the given logic values of the input signals, we performed measurements with variable concentrations of both inputs in these enzymatic systems. The output product was detected by optical and electrochemical means that gave close results when cast in terms of the logic variables.

From numerical analysis of the response surfaces we found that both logic gates possess similar noise characteristics, i.e., they significantly amplify random noise in inputs with a maximum noise amplification factor of 2.3 and 1.7 for the lactate/LDH and GSSG/GR gates, respectively. Both of them also exhibit systematic noise due to the non-zero product output at logic **0**. However, at realistic values of the input noise distribution widths, of the order of ~10%, it is nevertheless possible to distinguish the logic-**1** from logic-**0** outputs even though the three logic-**0** points **(0,1)**, **(1,0)**, **(0,0)** have substantial non-zero logic output values. This indicates that a reliable detection of abnormal physiological conditions[30] can be enabled by such logic gates. However, for larger spreads of the input noise or in a network of connected gates, noise suppression mechanisms, such as filtering, leading to a sigmoidal dependence of the output product on the chemical inputs, must be developed and utilized to allow for more complex information processing with biochemical logic.

**ACKNOWLEDGMENT:** This work was supported by the NSF (Grant CCF-0726698) and ONR (Grant N00014-08-1-1202).

# Figures and Schemes

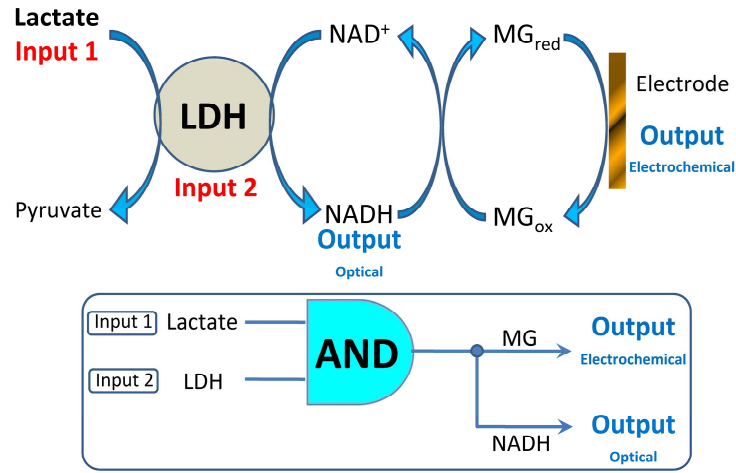

**Scheme 1.** The lactate/LDH biocatalytic cascade and its **AND** logic gate equivalent. $MG_{ox}$ and $MG_{red}$ are the oxidized and reduced forms of MG, respectively.

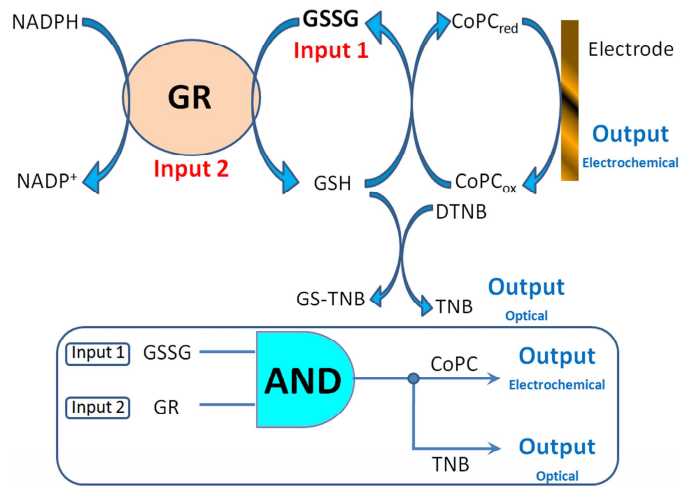

**Scheme 2.** The GSSG/GR biocatalytic cascade and its **AND** logic gate equivalent. $CoPC_{ox}$ and $CoPC_{red}$ are the oxidized and reduced forms of CoPC, respectively.

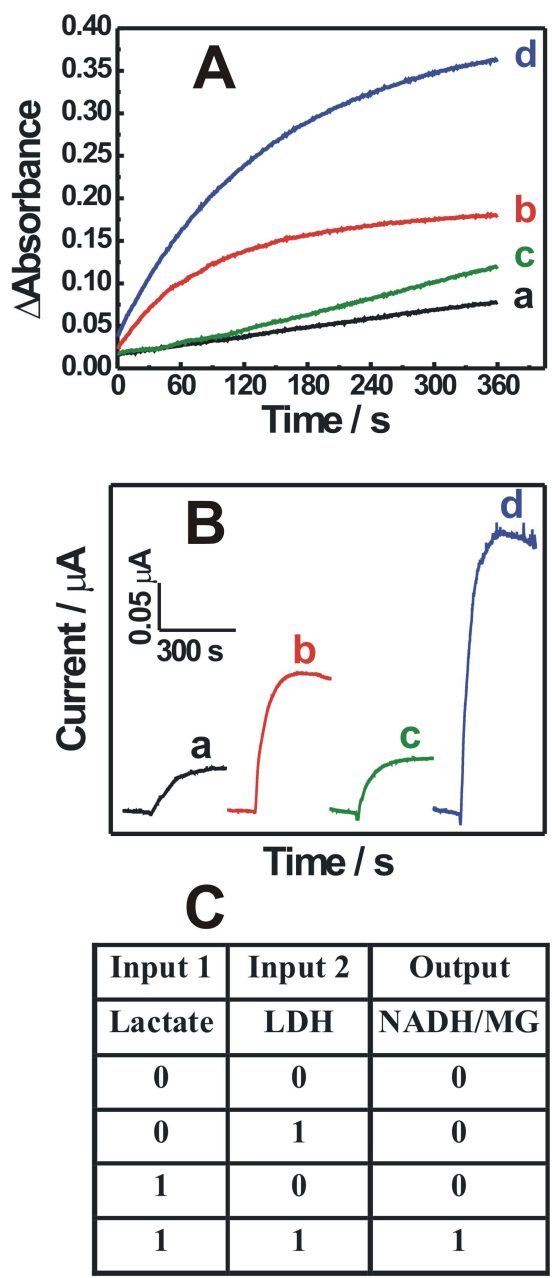

**Figure 1.** (A) Optical absorbance (λ = 340 nm), and (B) electrochemical amperometric (0.1 V) detection of the NADH output generated *in situ* by the lactate/LDH biocatalytic cascade upon different combinations of the input signals: a) **0,0**; b) **0,1**; c) **1,0** and d) **1,1**. The logic gate composition and the input concentrations corresponding to the logic **0** and **1** values are specified in the Experimental Section. (C) The truth table corresponding to the **AND** logic function of the system.

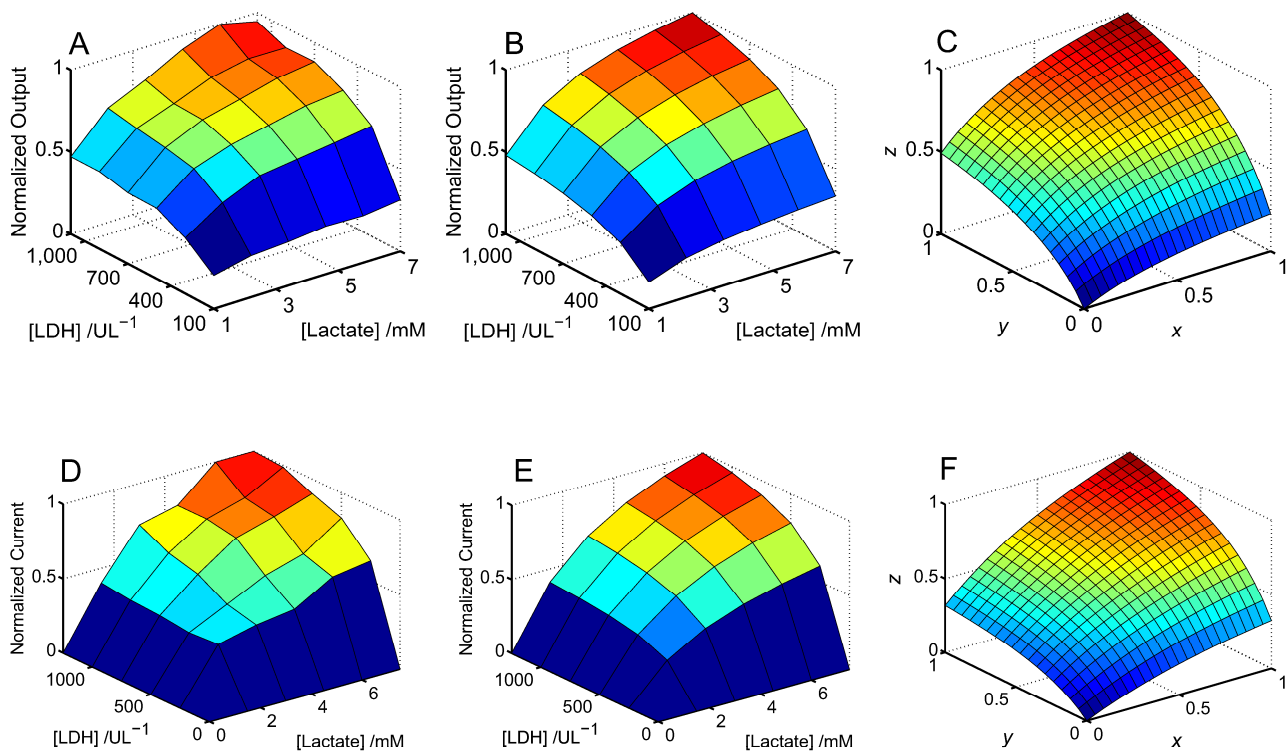

**Figure 2.** Experimental data and their analysis for (A)-(C) optical measurements, and (D)-(F) electrochemical measurements. Panels (A) and (D) show the experimental response surface for the LDH-based **AND** gate; (B) and (E) give the numerical fit according to Equation (3), with the resulting parameters $\alpha$ and $\beta$ found to be 0.645 (0.241) (mM)$^{-1}$ and $3.68 \cdot 10^{-3}$ ($4.10 \cdot 10^{-3}$) U$^{-1}$L, for optical (electrochemical) sets of data, respectively. Panels (C) and (F) show the response surface in terms of the logic-range variables $x$, $y$, $z$, properly scaled and *shifted* (since the logic **0** values are not defined at zero concentrations).

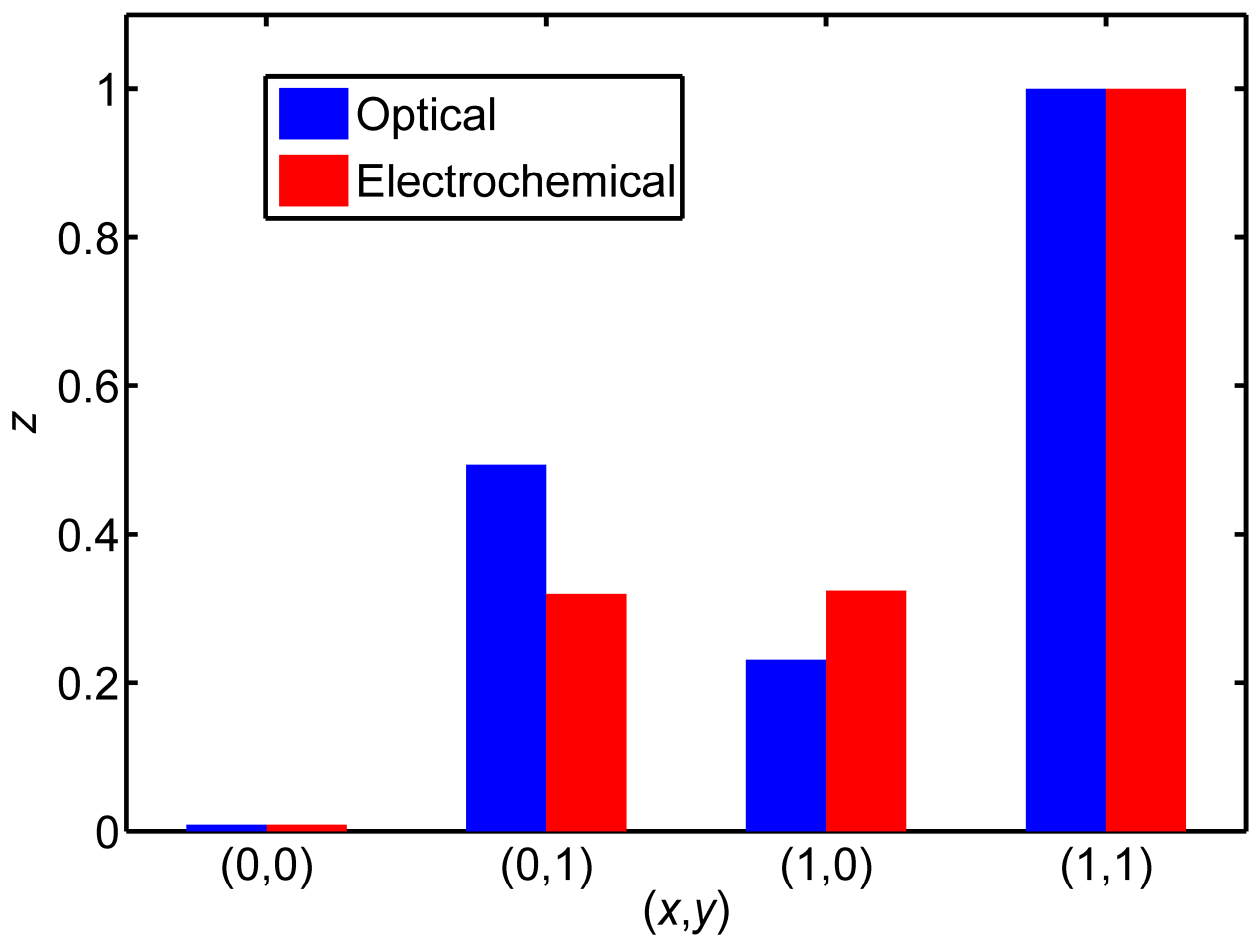

**Figure 3.** Normalized output logic values *z* for four logic inputs, for the different measurement techniques for the LDH-based **AND** gate. The logic gate composition and the input concentrations corresponding to the logic **0** and **1** values are specified in the Experimental Section.

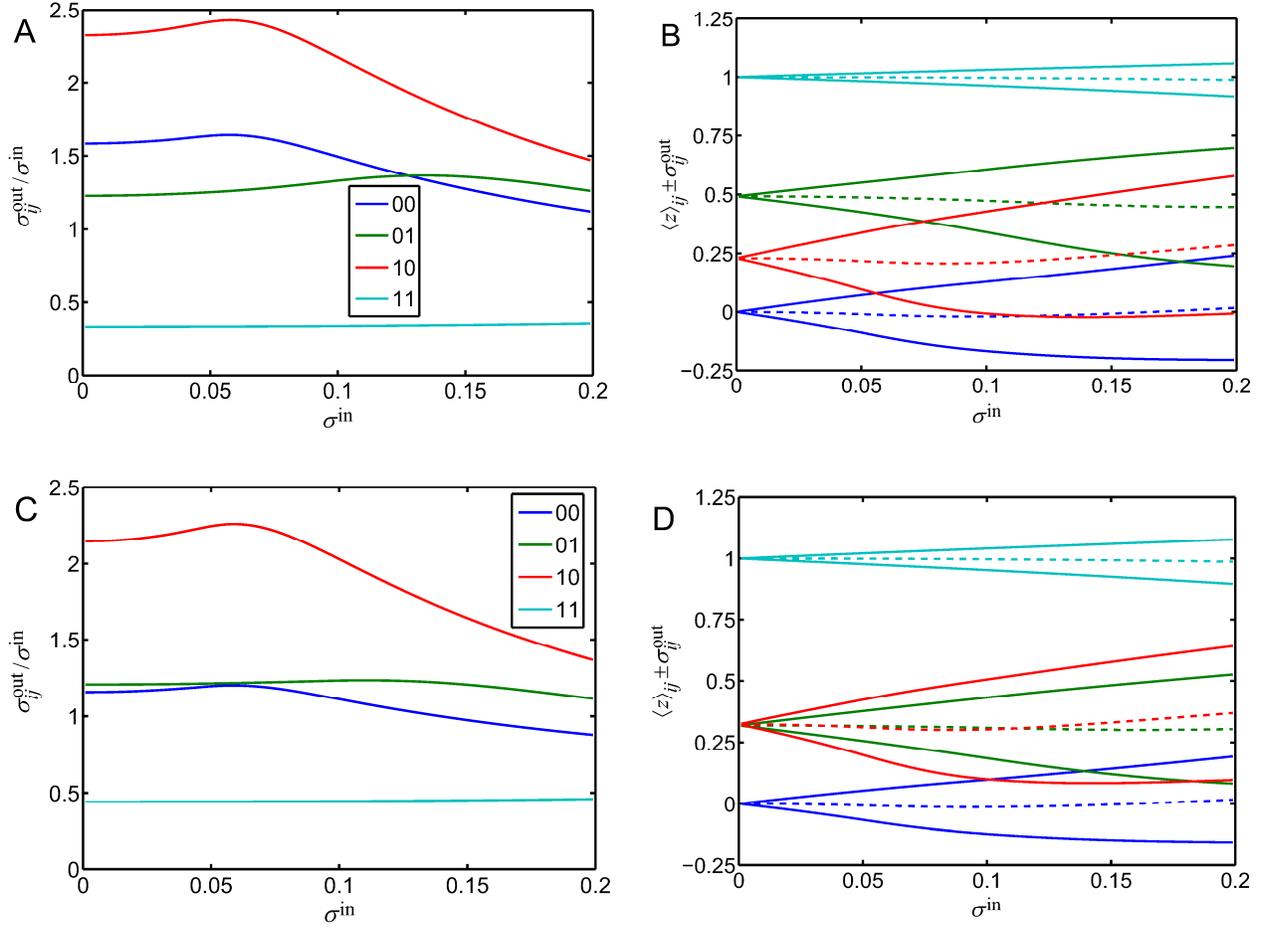

**Figure 4.** Noise propagation properties of the LDH-based **AND** gate, as calculated from fits of the optical, (A)-(B), and electrochemical, (C)-(D), data. Panels (A) and (C) show noise amplification factors $\sigma_{ij}^{out}/\sigma^{in}$ vs. the assumed width of the input noise distributions, $\sigma^{in}$. Panels (B) and (D) show the spread region vs. width of the input noise distribution $\sigma^{in}$. The dashed line is for the average value, $\langle z \rangle_{ij}$, of the logic output, while the solid lines of the same color give the upper and lower bounds of the spread region, $\langle z \rangle_{ij} \pm \sigma_{ij}^{out}$.

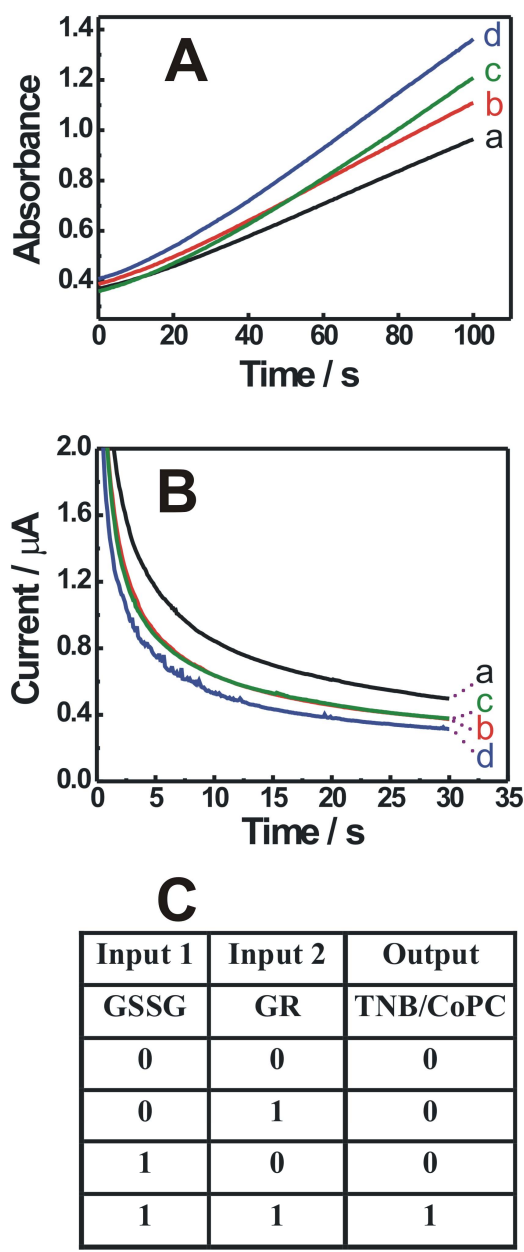

**Figure 5.** (A) Optical absorbance (λ = 412 nm), and (B) electrochemical chronoamperometric (0.5 V) detection of the GSH output generated *in situ* by the GSSG/GR biocatalytic cascade upon different combinations of the input signals: a) **0,0**; b) **0,1**; c) **1,0** and d) **1,1**. The logic gate composition and the input concentrations corresponding to the logic **0** and **1** values are specified in the Experimental Section. (C) The truth table corresponding to the **AND** logic operation of the system.

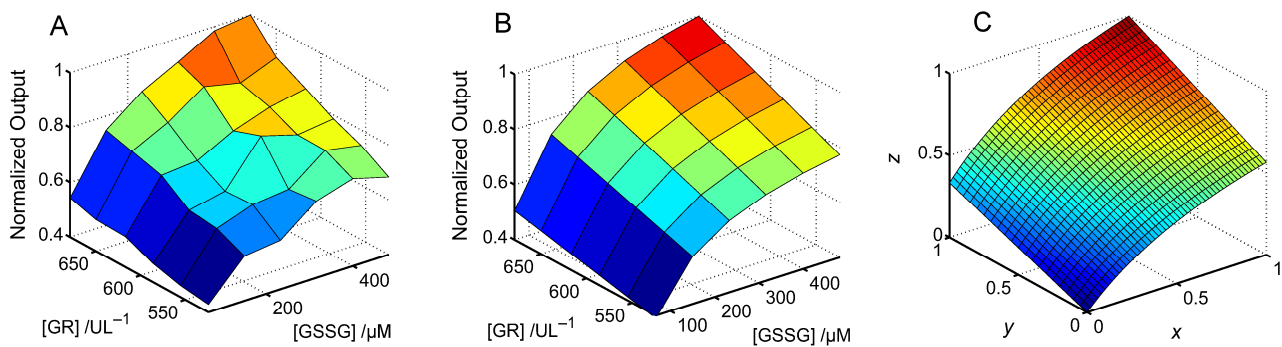

**Figure 6.** Optical measurements and numerical fitting of the response surface for the GR-based gate: (A) Experimental response surface. (B) Numerical fit of the surface in (A), according to Equation (3), yielding estimates $\alpha = 1.14 \cdot 10^{-2}\,(\mu M)^{-1}$ and $\beta = 0.00\,U^{-1}L$. (C) Logic surface $z(x, y)$.

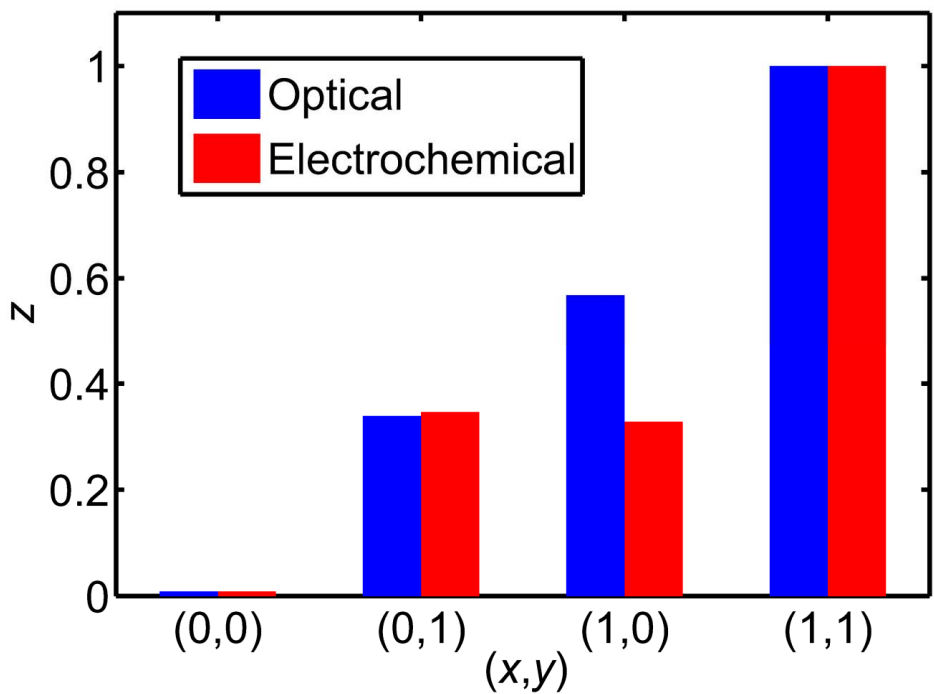

**Figure 7.** Normalized output logic values $z$ for four logic inputs, for the GR-based gate. The logic gate composition and the input concentrations corresponding to the logic **0** and **1** values are specified in the Experimental Section.

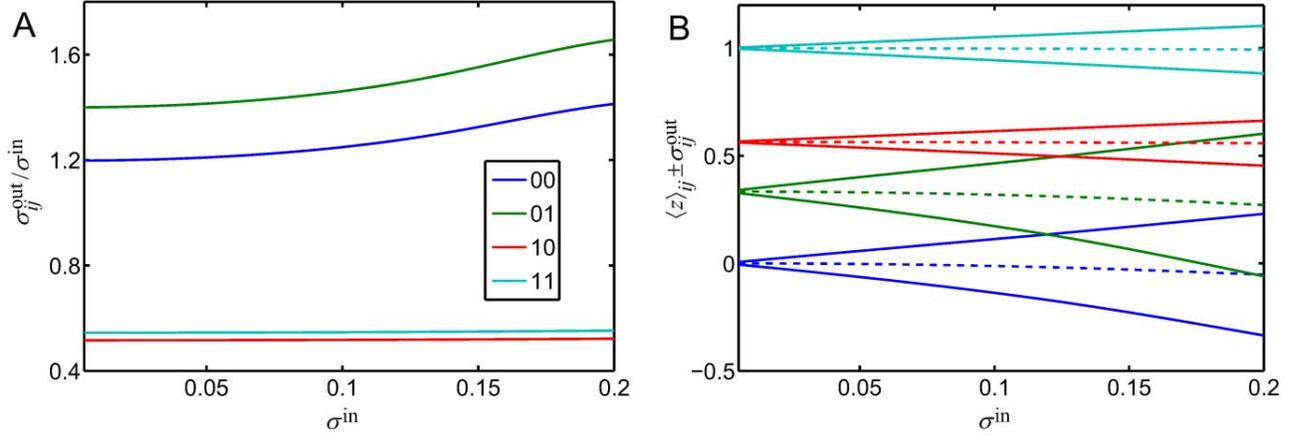

**Figure 8.** Noise propagation properties of the GR-based **AND** gate, as calculated from fit of the optical data. Panel (A) shows noise amplification factors $\sigma_{ij}^{\text{out}}/\sigma^{\text{in}}$ vs. the assumed width of the input noise distributions, $\sigma^{\text{in}}$. Panel (B) shows the spread region vs. width of the input noise distribution $\sigma^{\text{in}}$. The dashed line is for the average value of the logic output $\langle z \rangle_{ij}$, while the solid lines of the same color give the upper and lower bounds of the spread region, $\langle z \rangle_{ij} \pm \sigma_{ij}^{\text{out}}$.